\documentclass[graphicx]{emulateapj}

\slugcomment{}

\shorttitle{Spectroscopy of the ultra-compact binary V407 Vul}
\shortauthors{Steeghs et al.}

\begin{document}


\title{GEMINI spectroscopy of the ultra-compact binary candidate V407 Vul\footnote{This paper includes data gathered with the 6.5 meter Magellan Telescopes located at Las Campanas Observatory, Chile.}}


\author{D.Steeghs\altaffilmark{1}}
\affil{Harvard-Smithsonian Center for Astrophysics, 60 Garden Street, Cambridge, MA 02138, USA}
\author{T.R.Marsh, S.C.C.Barros\altaffilmark{2}}
\affil{Department of Physics, University of Warwick, Coventry CV4 7AL, UK}
\author{G.Nelemans, P.J.Groot, G.H.A.Roelofs\altaffilmark{3}}
\affil{Department of Astrophysics, Radboud University of Nijmegen, PO Box 9010, 6500 Nijmegen, The Netherlands}
\author{G.Ramsay, M.Cropper\altaffilmark{4}}
\affil{Mullard Space Science Laboratory, University College London, Holmbury St Mary, Dorking, Surrey RH5 6NT, UK}

\email{dsteeghs@cfa.harvard.edu}


\begin{abstract}

We present time-resolved spectroscopy of the optical counterpart to the proposed ultra-compact binary system V407 Vul (=RX J1914.4+2456). Our Gemini spectra resolve the 9.48 minute periodicity that has previously been reported for this source. We find that the optical counterpart is dominated by a reddened late-type spectrum of type G9V and contains solely unresolved absorption features. No radial velocity signatures exceeding 10 km/s could be detected on periods from minutes to hours. Using interstellar extinction estimates, we derive a distance to the G9 star of $1.1-3.5$ kpc. In addition to this stellar spectrum, we detect a blue component that modulates solely on the 9.48 minute period, and peaks $\sim 0.15$ in phase ahead of the X-ray peak. This blue component which contributes up to 40\% of the light shows no evidence for emission line features that are the usual hallmarks of an interacting binary. Good seeing images obtained with the Magellan telescopes indicate that the variable and the G-star are aligned to better than 0.1". Despite the low probability of a chance alignment of a field star along the line of sight, the G9 light cannot be directly associated with the 9.48 minute variable that powers the luminous ($\sim$10$^{35}$ erg/s) and highly variable X-ray source. The outlook for the detection of conclusive radial velocity measurements will remain challenging due to the extinction along the line of sight in conjunction with the contaminating effect of the G9V star.

\end{abstract}




\section{Introduction}

\begin{figure}[h]
\includegraphics[width=8.5cm]{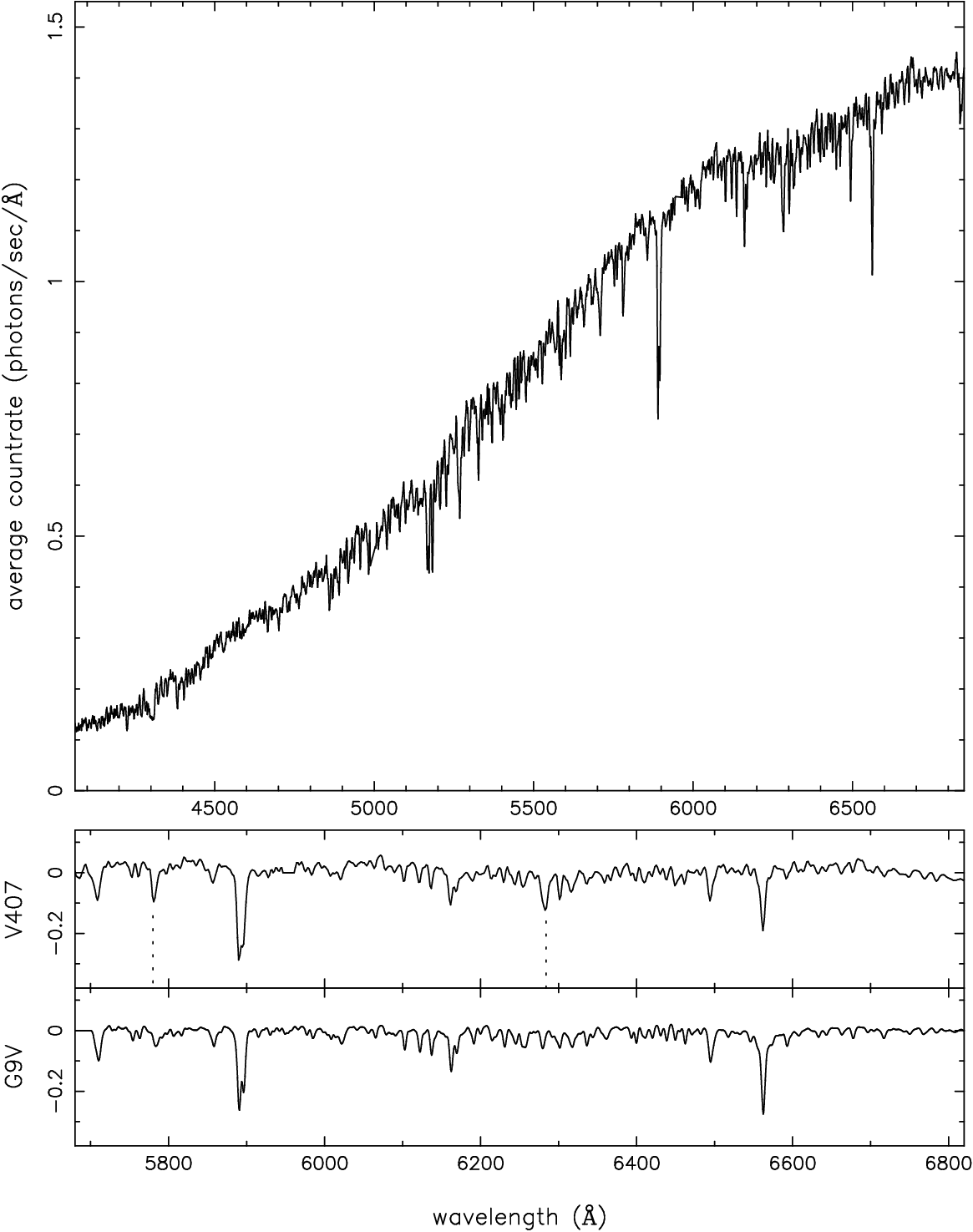}
\caption{Top: Average optical spectrum of V407 Vul revealing a reddened late G to
early K star absorption line spectrum.  Small gaps near 5000 and
5950\AA~are due to the gaps in between the individual CCD
detectors. The y-axis represents the mean counts s$^{-1}$\AA$^{-1}$~obtained in the 100s
exposures, no flux calibration was performed. Bottom panels compare
the spectrum of V407 Vul with that of a G9V star sampled at the same
resolution revealing a remarkable match. Dotted lines indicate the
location of two diffuse interstellar bands that reflect reddening
towards the source. \label{fig1}}
\end{figure}

Double white dwarf binaries form one of the largest populations of
compact binaries with a few hundred million systems expected across the
Galaxy (Napiwotzki et al. 2003). A substantial fraction of these are brought into contact at very short orbital periods (minutes) through
angular momentum loss via gravitational wave emission.  In the majority of cases, a rapid merger ensues due to unstable mass transfer (Webbink 1984, Iben 1990).
If the
combined mass of such double white dwarfs exceeds the Chandrasekhar
mass, a Type Ia supernova may be produced.  
A subset of such double white dwarfs are expected to survive the onset of
mass transfer and become stable semi-detached binaries where the more
massive white dwarf accretes from its companion via Roche-lobe
overflow. Such binaries are initially formed at orbital periods of a
few minutes and will evolve towards longer orbital periods as mass transfer drives
the two stars apart (Nelemans et al. 2001; Marsh, Nelemans \& Steeghs 2004).
Observationally a small group of such accreting white dwarfs are known, the so-called
AM CVn stars. These are hydrogen-deficient cataclysmic variables,
indicating that their mass donor stars are degenerate. Although
such systems can also be produced using a low-mass helium star donor (Savonije et al. 1986) or evolved post main-sequence donors (Podsiadlowski et al. 2002), the double white
dwarf formation channel is the only one that can produce systems with
orbital periods shorter than 10 minutes.

Recently, three variable stars  have been identified as candidate accreting
binary systems with ultra-short orbital periods; V407 Vul (Cropper et al.
1998; Ramsay et al.  2000), ES Cet (Warner \& Woudt 2002) and
RX J0806.3-0939 (Ramsay et al. 2002b; Israel et al.  2002).
Optical and X-ray light curves reveal large
amplitude variations on stable periods of 9.5, 10.3 and 5.3
{\it minutes} respectively, but on no other periods, suggesting that
these may be their orbital periods. 

V407 Vul was initially discovered as a {\it ROSAT} X-ray source
(RXJ1914.4+2456; Haberl \& Motch 1995), displaying a stable X-ray
periodicity of 569s which was suggested to represent the spin period
of an accreting magnetic white dwarf in an intermediate polar (IP, Motch et al. 1996). Cropper
et al. (1998) first suggested that V407 Vul might be a
double-degenerate binary with the X-ray period representing its
orbital period. This was motivated by the soft X-ray spectrum, the
shape of the X-ray lightcurve and the lack of other periodicities
which is very atypical for an IP. In this model, the accreting white
dwarf is magnetic and synchronised with the binary orbit. 
When the optical counterpart was identified (Ramsay et al. 2000), the
phasing of the optical modulation, the lack of strong emission lines
and the absence of polarised light raised new questions concerning the 
structure of V407 Vul. Several alternative models were suggested
which include a non-magnetic model where the accretion stream impacts
the primary directly due to the compact nature of double-degenerate
binaries (Marsh \& Steeghs 2002, Ramsay et al. 2002a), a unipolar
inductor model analogous to the Jupiter-Io system (Wu et al. 2002)
or a stream-fed IP viewed nearly face-on (Norton, Haswell \& Wynn
2004). Only in the IP interpretation would the 569s period not be the
orbital period but instead be the spin-period of the primary.

Strohmayer (2002,2004) analysed archival {\it ROSAT} and {\it ASCA} data as well as
new {\it Chandra} X-ray observations and found that the X-ray period was slowly
decreasing which appears at odds with an accreting ultra-compact
binary model since the period change is of the wrong sign. Although this appeared to support the non-mass transfer model of Wu et al. (2002), we may simply be seeing short term period changes that commonly occur in accreting binaries (Marsh \& Nelemans 2005).

In order to clarify the nature of V407 Vul and compare its
characteristics with the other AM CVn systems, we obtained high-time
resolution optical spectroscopy using the {\it Gemini-North} telescope that resolves the 9.48 minute 
period.
In Section 2, we discuss the observations and
data reduction and present the key observational results in Section
3.  In Section 4 we discuss the implications of our observations
concerning the nature of V407 Vul.

\section{Observations}

\subsection{Gemini spectroscopy}

We employed the {\it GMOS} spectrograph (Hook et al. 2004) mounted on the 8.1m Gemini-North
telescope on Mauna Kea, Hawaii and obtained time-series spectroscopy
of V407 Vul in queue observing mode. The detector array consisting of
three 2k$\times$4k CCDs was operated in $4\times$4 binning mode to
reduce readout overheads and sample the spectral resolution element
achieved with our 0.75'' slit across 2.3 binned pixels. The B600
grating delivered spectra covering 4058-6920\AA~at 1.8\AA/pixel, with
two small gaps near 5000\AA~and 5950\AA~at the interfaces between the individual CCDs. The slit was kept at a fixed
position angle in order to accommodate both V047 Vul and a nearby
comparison star in the slit. Weather conditions were non-photometric with a few exposures affected by thin cirrus while the seeing was between 0.75-0.8". Table 1 provides a log of observations.

\begin{table}
\begin{center}
\caption{Log of GEMINI observations}
\begin{tabular}{lccccc}
\tableline\tableline
UT Date & exp. & readout &\# exp. & UT interval & cycles \\
\tableline
Aug 9 2002 & 100s & 14s & 84 & 06:30 - 09:19 &  17.8 \\
Sep 2 2002 & 100s & 11s & 84 & 05:54 - 08:34 &  17.8 \\
Sep 4 2002 & 100s & 12s & 84 & 06:14 - 08:57 &  17.2 \\
\tableline
totals     &      &  & 252 & & 52.8 \\
\tableline\tableline
\
\end{tabular}
\end{center}
\end{table}

All frames were first de-biased using the overscan strips. Tungsten
exposures obtained at each observing block with the telescope pointed
at our target served to construct a flat-field frame that was used to
perform pixel-to-pixel sensitivity corrections.  The spectra of both
our primary target V407 Vul as well as the comparison star were then
optimally extracted from each frame.  Science frames
were bracketed by CuAr exposures that were used to wavelength
calibrate the spectra by interpolating the arc-scale derived from the
two nearest arcs.  Finally, the simultaneously recorded comparison
star spectra were used to correct the V407 Vul spectra for
time and wavelength dependent slit-loss effects by dividing the
individual comparison star spectra to the master comparison star
spectrum obtained at lowest air-mass. A second order polynomial was
then fitted to these ratio spectra and applied to the V407 Vul
spectra.  All relevant error contributions
were properly propagated throughout the reduction process to produce
final science spectra with accurate statistical errors. This reduction recipe delivered a 
total of 252 science spectra covering close to 53 cycles of the proposed 9.48 min orbital period.

\section{Analysis}

\subsection{The puzzling optical spectrum of V407 Vul}

The grand average optical spectrum obtained by averaging all 252 exposures with equal weights is shown in Figure 1. It reveals a reddened late-type stellar spectrum with a forest of
narrow absorption lines, including strong and narrow H$\alpha$ absorption. In line with the low S/N spectrum of Ramsay et al. (2002a), no emission line features can be discerned. 
The narrowness of the observed absorption lines, which we do
not resolve at our spectral resolution of 4\AA, is a surprise since an origin
in a binary system would broaden spectral features considerably in an
average spectrum due to binary motion. The optical spectrum of V407 Vul thus appears to be consistent with a normal stellar spectrum.
By comparing the average spectrum with late-type templates, we
estimate its spectral type to be close to G9/K0. Although we are somewhat limited in determining the luminosity class at our spectral resolution, the lack of giant features suggests a main-sequence classification. To illustrate, we compare the average spectrum to that of a G9V star in Figure 1 which indeed provides a very good match to all the observed features.
It is clear that it is not possible to fit a G9 star into a 9.48 minute binary system, and this identification thus intially appears in conflict with an ultra-compact binary scenario.

No emission features can
be discerned at the locations of strong hydrogen and helium lines in our average spectrum, nor is there evidence that the Balmer absorption lines are filled in by weak
emission. The signal to noise in our average spectrum, which
represents a total of 7 hours worth of exposure time, is around
125 per pixel at $H\alpha$. This is therefore large enough to be sensitive to
even very weak lines.
In all respects, the optical spectrum of V407 Vul resembles
that of a single star and appears at odds with the observed X-ray luminosity
and variability that has been reported.

\subsection{Time variability}

In order to look for the expected variability on the 9.48 minute period previously reported in X-rays and the optical, we constructed lightcurves by summing over our observed spectral range and calculated Scargle-normalised power spectra (Scargle 1982). These clearly reveal a coherent intensity 
modulation on the previously reported 9.48 min period with no significant signals detected at other periods. Although our uneven time-sampling limits a sensitive search for weak
side-bands, we do not find any evidence for them near the strong 9.48 min peak.
We thus find very similar optical variability behaviour as was discussed in Ramsay et al. (2002a).
The variability does confirm that we have acquired spectra of the correct optical counterpart, ruling out a possible mis-identification. A close inspection of the GMOS acquisition images near the slit position revealed no contaminating field stars that could have contributed to the spectra.

Although we detect a 5-15\% amplitude intensity modulation in our data, there appears to be no corresponding radial velocity motions. Cross-correlating the individual 100s exposures reveals
no significant radial velocity shift in any of the 252 spectra that
were obtained in three hour long blocks over a baseline of 26 days. 
Since our time-series cover of order 53 cycles of the 9.48 min period, we phase-folded our data on the
X-ray period using the ephemeris reported by Strohmayer (2004) in order to increase our sensitivity. The ephemeris was converted into the heliocentric frame and the UTC time-system and includes the measured period derivative term.

\begin{figure}[h]
\includegraphics[width=8.5cm]{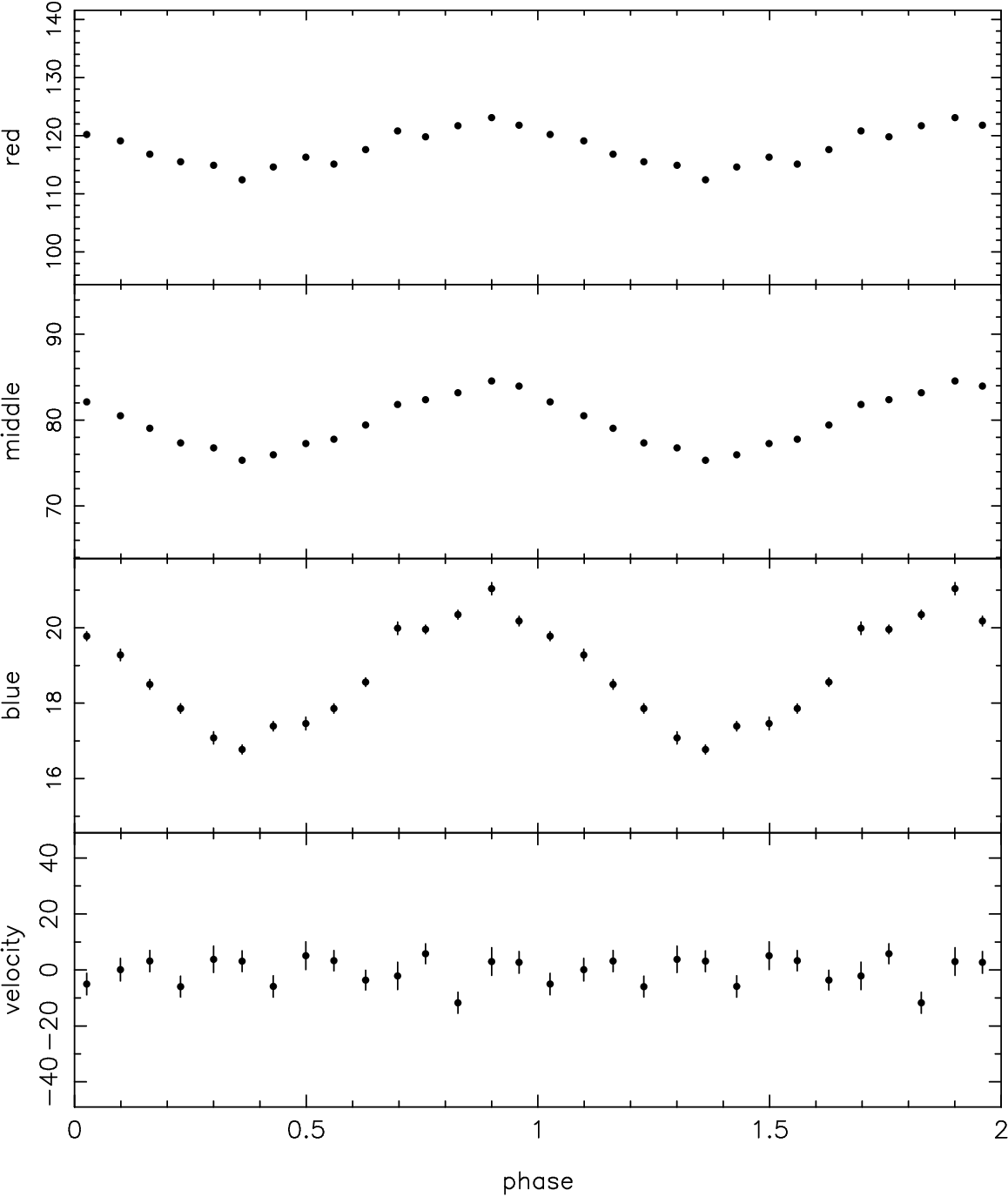}
\caption{Optical lightcurves after folding on the 9.48 min period using 15 phase bins. Two
cycles are plotted obtained in three wavelength bands: red (6200-6700\AA),
middle (5300-5800\AA) and blue (4100-4600\AA). The y-scale is adjusted
such that each panel spans $\pm$ 20 \% around the mean. Bottom panel
are radial velocity shifts derived by cross-correlating the absorption
line spectrum. \label{fig2}}
\end{figure}
We then constructed light-curves of various spectral segments after phase-folding,
revealing that the peak-to peak amplitude increases from 9\% in the red part of
our spectral range up to 25\% in the blue end (Figure 2). As remarked in previous studies, the optical peaks earlier in phase relative to the X-ray peak. In our case, this was 0.15 cycles in advance of the Strohmayer (2004) X-ray ephemeris.

In order to
quantify the lack of radial velocity motions in the absorption lines across the 9.48 minute period,
we also cross-correlated each phase-folded spectrum with the mean
spectrum. 
The derived radial velocities are also plotted in Figure 2
and reveal no motions phased with the 9.48 min period. All radial velocities are consistent with zero
at all phases. In order to place an upper limit on any possible radial
velocity amplitude, we shifted the spectra by various known amounts
and tried to recover the shifts using the above procedures. We found
that we can rule out any 'orbital' radial velocity shifts across the
9.48 minute period with amplitudes larger than $\sim$ 10 km s$^{-1}$. 

\subsection{Radial velocity limits on the period}
\begin{figure}
\includegraphics[width=6.5cm,angle=-90]{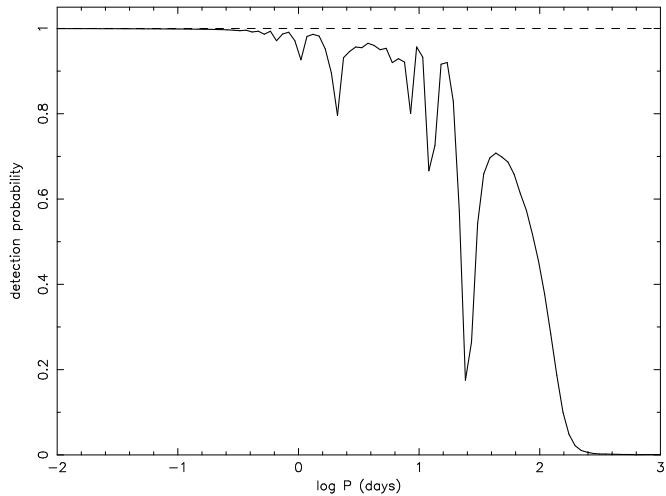}
\caption{Our detection probability for finding radial velocity motions as a function of the orbital period. The solid curve is re-binned to constant bins of 0.05 in $\log{P}$. Our detection probability is over 99\% for any periods below 14 hours. \label{fig3}}
\end{figure}
Our radial velocity time series revealed no detectable radial velocity motions in the G-star spectra whatsoever. We ran a period detection algorithm in order to evaluate our detection limits on any periodic motions in the G-star. 
%
We define a detection as occurring when the $\chi^2$ for a fit of a 
constant to the velocities is sufficiently large as to have a very small 
probability $p$ of occurring by chance (we took $p = 10^{-3}$). For a 
given orbital period, we then calculate the orbital speed of the G star
induced by the variable taking the masses to be $\sim0.8 M_{\odot}$  for the G-star and 
$\sim0.6 M_{\odot}$ for the white dwarf primary. 
Orbital velocities for a 9.48 minute binary are of order 1000 km s$^{-1}$, but even for periods as long as 10 hours would still be larger than 100 km s$^{-1}$.
This translates into a lower limit on the
orbital inclination $i_c$ which will trigger a detection. Assuming 
randomly inclined orbits, the detection probability is then $cos(i_c)$.
Averaging this over the zeropoint of the orbital phase assuming it is 
uniformly distributed from 0 to 1 gives us the overall detection 
probability for a particular orbital period.  
The observational uncertainties in the velocities are taken into account through Monte Carlo generation of orbital velocities plus noise,
in the form of Gaussian white noise. 
Finally, we re-binned the resulting detection probabilities into period bins spanning 0.05 in $\log{P}$ for periods ranging between a few minutes and 100 days.  These detection probabilities are plotted in Fig.3.
Our detection probability is over 99.9\% for any periods under $\sim3$ hours and remains better than 99\% for periods up to $\sim14$ hours.  This thus argues against identifying the G9-type star that dominates the optical light as the donor star in a compact binary system.
We return to the implications of these constraints in Section 4.

\subsection{Decomposing the variable part}
\begin{figure}
\includegraphics[width=6.5cm,angle=-90]{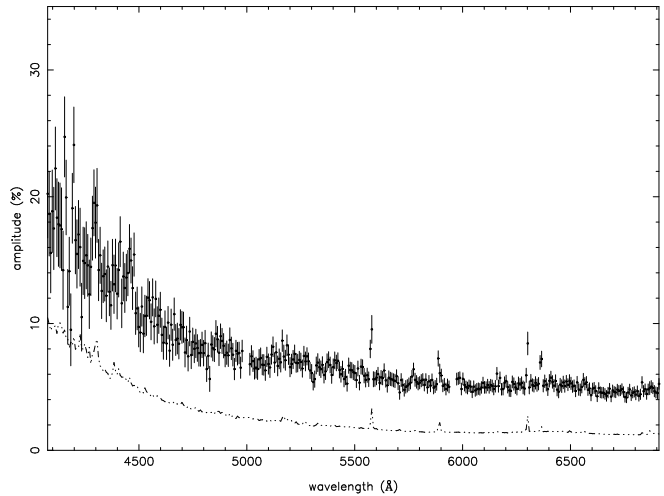}
\caption{Wavelength dependence of the 9.48 min modulation after binning to 7\AA. The fitted value is the semi-amplitude of a sine wave in phase with the modulation. The dashed curve denotes our 3$\sigma$ sensitivity limit to the semi-amplitude at each wavelength.\label{pdetect}}
\end{figure}

In order to decompose the spectrum and determine the spectral
characteristics of the variable part of the light, we fitted a
sine-curve to the time-series of each wavelength pixel while fixing
the period to the known X-ray period of 9.48 min. We thus derived the
semi-amplitude of the modulation as a function of wavelength. This
modulation spectrum is plotted in Fig.4 and shows a relatively flat
modulation semi-amplitude of 5\% in the red, rising steeply and approaching 20\% on the
blue end, in line with the modulations displayed by the folded
lightcurves in Fig.2. No spectral features, either in absorption or emission, are detected in this
modulation spectrum with significance, apart from a few bright pixels
that correspond to a small amount of contamination due to strong sky
lines. In particular, near the location of potential Hydrogen and Helium features, we see no evidence for any excess variability. It thus appears that also the variable part of the light shows no evidence for emission lines that are the hallmarks of an accreting binary. Although our sensitivity to weak lines in the modulated spectrum is obviously worse than our ability to see them in the average spectra, we still have the sensitivity to pick up emission lines at the strengths observed in typical cataclysmic variables. For example, at H$\alpha$, we have the sensitivity to pick up emission lines with equivalent widths larger than $\sim$ 4\AA ($5\sigma$ limit).

The above decomposition of a non-varying G-star spectrum plus a variable blue component appears to describe our time-series fully. We checked whether there was any evidence for a non-variable component in addition to the two components described above by studying the depth of the G-star absorption lines relative to the continuum. Any additional light source would dilute the absorption lines relative to that of an uncontaminated G-star. We find that this dilution is at most 10-15\% at the red end of our spectral range, rising to 30-40\% at the very blue end. This is consistent with the full amplitude of the variable component and indicates that any additional contribution to the light is rather small.

\subsection{Reddening and distance}

The line of sight toward V407 Vul suffers from considerable
extinction although the estimates vary significantly. According to the dust maps of Schlegel et al. (1998), the integrated line of sight column is $N_H = 3.2 \times 10^{21}$ cm$^{-2}$ using the $N_H/E(B-V)$ scaling as derived by Predehl \& Schmitt (1995). Joshi (2005) derived a mean extinction of $A_{V} \sim 1.5$ mag kpc$^{-1}$ ($\sim 2.7 \times 10^{21}$ cm$^{-2}$ kpc$^{-1}$) in the region towards V407 Vul though with considerable local scatter.
Spectral fits to the X-ray spectra from various missions have also lead to a range of derived column densities.
 Haberl \& Motch (1995) determined a column of $N_H = 1.0
\times 10^{22}$ cm$^{-2}$ using an absorbed black-body fit to the {\it ROSAT}
spectrum, although Ramsay et al. (2000,2002a) remark that the $N_H$ is not closely constrained from fits to either {\it ASCA} or {\it ROSAT} data.
More recently, Ramsay et al. (2005) find  $N_H = 4.2
\times 10^{21}$ cm$^{-2}$ from fits to {\it XMM-Newton} data, and a re-analysis of the same data in Ramsay, Cropper \& Hakala (2006) gives an even lower column of $N_H = 1.2 \times 10^{21}$ cm$^{-2}$ when applying a different plasma model. This range in reported extinction spans nearly an order of magnitude and has considerable impact on the derived X-ray luminosities, which range from $3 \times 10^{31}$ to $10^{35}$ ergs s$^{-1}$.

In our data, a substantial amount of reddening is apparent through the
steep slope of the optical spectrum shown in Fig.1. No flux standard data was
available that would permit us to calibrate the Gemini
spectra. However, our wavelength range includes several diffuse
interstellar bands that can serve as reddening indicators (DIBs; Fig 1).
The ratio between the two components in the prominent Na-D doublet
near 5895\AA~was close to 1, indicating saturation. This doublet thus can only provide a lower limit of $E(B-V)>0.5$ ($N_H > 3 \times 10^{21}$ cm$^{-2}$). However, the 5780\AA~DIB is present
with an $EW(5780)=0.7 \pm 0.1$\AA. Using the $E(B-V)$ vs $EW$ relation as
published in Herbig (1993), this translates to a $E(B-V) = 1.4 \pm 0.2$ ($N_H \sim 7 \pm 1 \times 10^{21} $cm$^{-2}$) 
towards V407 Vul. These values are subject to the accuracy of the calibrations used, for which significant variations along specific lines of sight are possible.
The DIB features in our mean spectrum thus suggests that the reddening towards the G-star is substantial.
We also determined the strengths of these DIBs in the comparison star that was present in our slit and find equivalent widths consistent with those of V407 Vul.

As a final consistency check, we tried to reproduce the observed broad-band colors of V407 Vul as listed in Ramsay et al. (2002a). We were not able to fit all the colors satisfactorily with a model consisting of a reddened G9-star plus a blue component. Particularly puzzling are the near-infrared colors, which should be dominated by the G9-star yet cannot be described by reddened stellar models. A reasonable match to the optical colors can be achieved if we decompose the colors into a reddened G9-star with $N_H \sim 3-4 \times 10^{21}$ cm$^{-2}$ plus a blue component with contributions in line with Fig.4. Models with reddening values as large as those inferred from the 5780\AA~DIB would overpredict the contribution from the variable in the blue bands considerably.
%

%
%

Given V407 Vul's observed $R$ magnitude of $R=19.2 \pm 0.1$ (Ramsay et al. 2002a), this leads to a de-reddened $R$-band magnitude of $15.5-17.9$ using the Schlegel et al. (1998) extinction scaling and our derived range of $N_H=3-7 \times 10^{21}$ cm$^{-2}$.  Since the G9 star light totally dominates the R-band, we can estimate the distance to this star using a G9 absolute magnitude of $M_R(G9V) \sim 5.2 \pm 0.3$  (Houk et al. 1997) and find $d= 1.1-3.5$ kpc. We remark that the uncertainty in the derived distance is completely dominated by the uncertainty in the assumed $N_H$.
%
 Since we cannot firmly rule out a (sub)giant classification, these are strict lower limits on the distance. According to Neckel \& Klare (1980), the interstellar absorption ramps up rapidly within 1kpc in this region and our distance and hydrogen column from the DIBs are thus in that sense consistent, while it would be puzzling if the distance was much larger yet the reddening much lower. 


Whether this represents the distance to the variable optical/X-ray
source is not certain since the observed properties of the G9 star
make it impossible to associate it directly with either the X-ray source or
a member in a compact binary. Our interstellar bands sample the column towards the G star, and the apparent disparity with the recent X-ray inferred columns could indicate that the X-ray source is not at the same distance as the G9 star, but significantly closer. The observed X-ray flux shows significant variability (Ramsay et al. 2000,2005,2006, Strohmayer 2004).  If we assume that the X-ray source is at the same distance as the G9-star, the inferred bolometric X-ray luminosity is $10^{33}-10^{35}$ erg s$^{-1}$ depending on the spectral model used. This is clearly far beyond any chromospheric X-ray emission that could be powered by a G9 star, let alone be modulated at 9.48 minutes.
%
%

\subsection{Chance alignment constraints}

Given the puzzling properties of the G9 starlight that dominates the optical spectrum of such a strong and variable X-ray source, we investigated the possibility whether the G9 star is merely a chance alignment along the line of sight, and is not actually associated with the 9.48 minute variable.

On June 23 2003, we obtained $BVRI$ images of the field surrounding V407 Vul with the MAGIC camera on the Magellan-Clay telescope at Las Campanas Observatory. The pixel scale was 0.069" and seeing conditions were very good, achieving a point source FWHM = 0.45". The optical counterpart to V407 Vul was detected as a non-resolved point source in all images. We also compared the position of the counterpart relative to nearby field stars and found that these positions matched to better than 1 pixel when comparing B,V,R and I-band frames. We therefore conclude that the source dominating in the I-band (the G9 star), and the variable source that contributes $\sim$ 30\% to the light in the B-band are aligned to within 0.1".
By counting all stars in our imaging data down to the brightness of V407 Vul, we derive a local field star density of $4.6 \times 10^{-3}$ stars arcsec$^{-2}$. The probability of chance alignment of a field star at the level of $<0.1$" as derived from our imaging analysis amounts to $\le 1.4 \times 10^{-4}$. 
These numbers are consistent with the even tighter alignment constraints derived in Barros et al. (2006, in prep) based on high-speed optical photometry of V407 Vul.
While the lack of emission lines or radial velocity motions argues against relating the G9 dominated spectrum directly with the source of the variable light, this small probability suggests that a physical connection needs to be considered.

\section{Discussion}

Our phase-resolved spectroscopy reveals that the optical light of V407 Vul is dominated by a G9V-type stellar spectrum, viewed through a 
column of $N_H= 3-7 \times 10^{21}$ cm$^{-2}$ and at a distance of $d= 1.1-3.5$ kpc. Good seeing images indicate that this G9 star is very closely aligned with the 9.48 minute variable source making a chance alignment with a field star unlikely. However, the complete lack of radial velocity signatures on periods from minutes to hours and the absence of any emission features appears to rule out that this G9 star is a member of a compact binary that powers a $10^{35}$ erg s$^{-1}$ variable X-ray source.
Although the G9 totally dominates our spectra in the red end, the broad-band colors of V407 Vul cannot be modeled with a simple reddened G9 star model. Ramsay et al. (2000,2002a) also discuss this issue and find that the infrared colors are not consistent with reddened late-type stars. However, the magnitudes in the various colors were not obtained simultaneously and Ramsay et al. (2000) report a change in the K-band magnitude of 1.1 mag between two epochs. Such significant variability and the apparent mismatch of colors in the IR presents yet another puzzling aspect that does not fit obviously with the apparently non-varying G9 starlight that dominates our spectra. Perhaps another spectral component starts to contribute towards the red. A contemporaneous study of the optical/IR colors and variability appears to be warranted so that these characteristics can be reconciled.

\subsection{A triple system?}
A possibility could be that the G9 star is part of a triple system, consisting of the G9 star in a relatively wide orbit around the compact binary that is responsible for the observed variability and X-ray emission. This would lead to no obvious radial velocities on short periods, nor would there be emission lines since the G9 star is relatively far removed from the compact binary. 
Although bringing in their own formation and stability concerns, triple star systems are relatively abundant and the low mass X-ray binary V1727 Cyg, for example, is thought to be a hierarchical triple (Garcia et al. 1989).
An interesting consequence of a triple scenario is that the observed period derivative of V407 Vul (Strohmayer 2002, 2004, Ramsay et al. 2005,2006) may then be influenced or even dominated by orbital time-delay effects due to the G9 star. In that case, the observed period derivative cannot be interpreted as tracking the secular mass transfer rate of the ultra-compact binary. This makes the period derivative a less clear-cut test of the accretion geometry and the expected angular momentum loss rate from gravitational wave emission (see also Marsh \& Nelemans 2005).
If the observed period drift instead tracks the time-delays due to a much longer and wider orbit, we would expect cyclical period changes modulated with this long orbital period. Given that the current period drift of $\sim$400s over $\sim$4,000 days shows no sign yet of turning over (e.g Ramsay et al. 2006) we have only covered a modest part of the orbit during the 11 years of period studies. This in turn implies that the orbital period of the G9 star in this scenario would be larger than $\sim$25 years or so, well beyond what is needed for dynamical stability.

Alternatively, despite the low probability, the G9 may be physically unrelated to the 9.48 minute variable but happens to be extremely well aligned. 
The X-ray source could then indeed be much closer, which would alleviate the issue of the low hydrogen columns that are inferred from the X-ray spectra.
Either way, it is clear that the G9 star cannot be associated directly with the intriguing variable source that makes V407 Vul so fascinating. It unfortunately severely inhibits our ability to study the properties of this enigmatic variable. We remark that this also implies that care must be taken when interpreting observed magnitudes and modulation amplitudes since the observed light is a composite between the G9 star and the blue variable.
Despite the dominance of the G9 light, we were able to clearly detect a blue component that modulates coherently on the previously reported 9.48 minute period, with no evidence of variability at other periods sampled by several efforts at different wave-bands to date. This component starts to make an appreciable contribution in the blue end of our spectral range and it appears that despite the significant reddening along the line of sight, it is at the blue end that we may be able to constrain the properties of V407 Vul. As discussed in Section 3.3., even this variable component to the light shows no evidence for emission line features that would be expected for a mass transferring compact binary. 
If, as the most recent X-ray analysis may suggest, the X-ray source actually suffers from much less reddening compared to the G9 star, deeper UV studies are feasible and could provide important constraints on the bolometric flux from the variable.

Since the discovery of V407 Vul as a variable X-ray source, a number of models have been put forward for V407 Vul and its related short period cousins RXJ0806 and ES Cet. We do not wish to repeat all the pros and cons of the various models in detail, but will focus on those features relevant for the spectroscopy results presented here.

\subsection{A stream-fed IP?}

Most scenarios identify the sole periodicity that has been detected in these systems as the orbital period of an ultra-compact binary. An alternative model put forward by Norton et al. (2004) suggests that these periods represent the beat period between spin and orbit of a magnetic white dwarf viewed at very low inclinations and accreting from a stream instead of an accretion disc (a so called stream-fed IP).  The orbital period of the binary in the stream-fed IP scenario would be 1.5-2.5 hours for V407 Vul, which would lead to orbital velocities of $>$ 250 km s$^{-1}$. Our period analysis in Section 3 also illustrates that our detection probability for periods in the IP range is very high, yet we see no evidence for orbital motion. Given our measurement limits, the inclination would then have to be less than two degrees to explain the lack of measurable radial velocities.
An additional issue is the total lack of emission features. Although Norton et al. argue that a narrow stream feeding the WD may be optically thick and this would not produce emission lines, this is not what is observed in other accreting systems. Despite the presence of optically thick disks in non-magnetic CVs and the presence of dense, optically thick streams in magnetic polars and IPs, all these systems invariably shows very strong emission lines from optically thin regions above those disks and streams. The situation would be very similar to that of high-state polars where accretion stream emission is visible out to the L1 point (e.g. Schwope et al., 1997). Indeed, the prototypical stream-fed IP V4200 Oph shows very strong hydrogen and helium emission features (Hellier \& Beardmore 2002). We have the sensitivity to see such strong lines both in the grand average spectrum (Fig.1) as well as the variable part of the light (Fig.3) and see no such lines in either. In summary, we see no evidence for the two tell-tale signs that would be expected for a low-inclination IP (radial velocities on $P_{orb}$ and emission lines). The continued lack of evidence for additional periodicities in the now rather extensive number of observations at different wavebands and the atypical X-ray properties of V407 Vul are additional issues which reduce the appeal of an IP interpretation.

\subsection{An ultra-compact binary hiding behind a star?}

All remaining scenarios have identified 9.48 minutes as the orbital period of the binary. As argued above, this would imply that the G9 star is not directly related to the ultra-compact binary and merely dilutes and complicates the analysis of the underlying variable.

In the double-degenerate polar model of Cropper et al. (1998) and the direct-impact model (Nelemans et al. 2001, Marsh \& Steeghs 2002, Ramsay et al. 2002a) the X-rays are powered by mass transfer onto the primary white dwarf. One would thus expect to see emission lines from the accretion flow in those cases. In the unipolar-inductor model of Wu et al. (2002), there would be no mass transfer and thus no emission lines from the accretion flow itself, though there could be emission from the irradiated mass donor star.
The fact that the variable part of the light is very blue suggests an origin in a hot environment, which in the accreting ultra-compact models could naturally be associated with the accreting primary that is expected to be heated to temperatures in excess of 50,000K (Bildsten et al. 2006). 

Since our data resolve the 9.48 min period, we would also be able to pick up rapidly moving narrow components that would be expected if the emission lines originate along a stream or on the surface of the rapidly rotating donor star. We certainly do not see the kind of strong helium emission lines that are typically observed in the longer period disk-accreting AM CVn systems (e.g. Nather, Robinson \& Stover 1981, Roelofs et al. 2005), including the 10.3minute system ES Ceti (Wegner et al. 1987; Warner \& Woudt 2002). However, a very weak emission line spectrum, such as is observed in RX J0806.3-0939 (Israel et al. 2002) could actually be hidden in the seemingly featureless spectrum of the variable source (Fig.3). While our signal to noise is good enough to see lines of typical strengths such as displayed in all the other CVs and AM CVns, very weak lines would go undetected. The fact that such short-period systems may not accrete via a disk but instead via a (direct impact) stream could be one of the reasons for a reduced emission line contribution. 
The detection of emission line components showing radial velocity motions across the 9.48 min orbit would have been conclusive proof of the binary period, and indeed was the aim of our experiment. We are thus severely hampered by the diluting effect of the G9 starlight in conjunction with the reddening along the line of sight to make conclusive statements concerning the presence or absence of very weak helium lines. A significant improvement will be difficult to make given the faintness of the source at blue wavelengths.

\subsection{Outlook}

The true nature of V407 Vul continues to puzzle us. The search for radial velocity signatures of its orbit remains elusive. Additional, high-resolution, spectroscopy in the red end will permit more stringent constraints to be placed on the properties of the G9 star that dominates the light. If this star is indeed in a longer orbit, accurate velocity measurements sampling longer base-lines could for example confirm the possibility of a triple. While the underlying variable is best studied in the blue, the reddening and intrinsic faintness of the source makes this a rather challenging avenue. After all, we were not able to detect features even with a considerable amount of 8m-class spectroscopy. However a rigorous analysis of the spectral energy distribution of the source including sensitive UV flux constraints should shed more light on the relative contributions of the G9 star and the variable powering the X-ray source.

\acknowledgments

DS acknowledges a Smithsonian Astrophysical Observatory Clay Fellowship as well as support through NASA GO grants NNG04GG30G and NNG04GG96G. GN is supported by NWO-VENI grant 639.041.405. TRM acknowledges the support of a PPARC
Senior Research Fellowship. SCCB is supported by FCT.

We thank Josh Grindlay for usefull comments related to the triple scenario.
We would like to thank the Gemini staff for their help in the planning
and acquisition of the spectra discussed in this paper. 
This work uses observations obtained at the Gemini Observatory, which is operated by the Association of Universities for Research in Astronomy, Inc., under a cooperative agreement with the NSF on behalf of the Gemini partnership: the National Science Foundation (United States), the Particle Physics and Astronomy Research Council (United Kingdom), the National Research Council (Canada), CONICYT (Chile), the Australian Research Council (Australia), CNPq (Brazil), and CONICET (Argentina).
This work uses data obtained with the Magellan telescopes. The Magellan project is operated by Carnegie Observatories at Las Campanas Observatory, Chile on behalf of the Magellan consortium. 








\end{document}